# Unintentional Consequences:
# Generative AI Use for Cybercrime

Truong (Jack) Luu
University of Cincinnati, Cincinnati, United States
luutp@mail.uc.edu

Binny M. Samuel
University of Cincinnati, Cincinnati, United States
samuelby@uc.edu

## Abstract

The democratization of generative AI introduces new forms of human-AI interaction and raises urgent safety, ethical, and cybersecurity concerns. We develop a socio-technical explanation for how generative AI enables and scales cybercrime. Drawing on affordance theory and technological amplification, we argue that generative AI systems create new action possibilities for cybercriminals and magnify pre-existing malicious intent by lowering expertise barriers and increasing attack efficiency. To illustrate this framework, we conduct interrupted time series analyses of two large datasets: (1) 464,190,074 malicious IP address reports from AbuseIPDB, and (2) 281,115 cryptocurrency scam reports from Chainabuse. Using November 30, 2022, as a high-salience public-access shock, we estimate the counterfactual trajectory of reported cyber abuse absent the release, providing an early-warning impact assessment of a general-purpose AI technology. Across both datasets, we observe statistically significant post-intervention increases in reported malicious activity, including an immediate increase of over 1.12 million weekly malicious IP reports and about 722 weekly cryptocurrency scam reports, with sustained growth in the latter. We discuss implications for AI governance, platform-level regulation, and cyber resilience, emphasizing the need for multi-layer socio-technical strategies that help key stakeholders maximize AI's benefits while mitigating its growing cybercrime risks.

## 1. Introduction

An accounts payable employee receives a call. The voice on the other end is their boss, the tone familiar, the request urgent: transfer thousands of dollars to settle an unexpected invoice. The employee hesitates, knowing it's against protocol, but the voice is persuasive, the pressure is real, and the request seems plausible. The transfer is made. The problem? The "boss" was never on the phone. It was a deepfake, an audio illusion crafted by generative AI, and the money is gone forever (Renaud et al., 2023). This cyber risk is consistent with the growing technical emphasis on scalable deepfake detection methods as synthetic media becomes easier to produce and harder to verify (Saif et al., 2024).

The rapid democratization of powerful generative AI models like ChatGPT has been celebrated for its potential to revolutionize productivity and creativity. Yet, this leap casts a long shadow. The very qualities that make generative AI a remarkable tool for innovation, its ease of use, speed, and ability to generate human-like content, also make it a suitable instrument for those with malicious intent (Leong et al., 2025). As one headline warns, our natural *'spidey sense'* is no match for the new wave of scammers (Brown & Hamilton, 2024). In parallel, scholarship highlights how generative systems reshape information flows and trust during crises, intensifying the harm from convincing synthetic content (Kim et al., 2023; Shahbazi & Bunker, 2024; Zhong, 2023). While researchers and security experts have issued stark warnings about AI's potential to scale cybercrime, from creating hyper-realistic phishing emails to writing malware with minimal expertise (Bengio et al., 2024), much of the evidence has been anecdotal or speculative. The critical question is no longer if GenAI can be misused for cyberattacks, but *to what extent* it is already happening in practice, *through which mechanisms*.

To answer this question, we adopt a socio-technical framework grounded in human-computer interaction, combining affordance theory and the concept of technological amplification to explain how formally neutral AI tools can scale malevolent human intentions. We then empirically illustrate this framework with interrupted time series analyses of two large cybercrime-related datasets: A general dataset of 464,190,074 malicious IP address reports from AbuseIPDB and a specialized dataset of 281,115 cryptocurrency scam reports from Chainabuse.

Using the public release of ChatGPT 3.0 on November 30, 2022 as a pragmatic intervention date, we analyze trends before and after this event to provide measurable evidence of how the public availability of GenAI relates to cybercrime-related behaviors. Our contributions are threefold. First, we theorize how GenAI capability releases create unwanted affordances that amplify malicious intent at scale. Second, we provide an early impact evaluation using interrupted time series models that quantify level and trend shifts in two large cybercrime reporting streams. Third, we translate the findings into a concrete governance and monitoring agenda spanning developers, platforms, and regulators. Positioning generative AI as a general-purpose technology, we evaluate its early societal consequences in the form of measurable shifts in cyber abuse and cryptocurrency scam reporting. We further translate these findings into actionable considerations for technology governance and risk management for dual-use AI systems.

## 2. Conceptual Background

Generative AI (GenAI) models were primarily designed to augment human productivity and enhance efficiency. However, as with any technology, there exists an inherent risk of misuse. The need to understand the unique capabilities/characteristics of novel tools first leads to a discussion of affordance theory, which focuses on understanding potential impacts based on the unique capabilities of a tool.

## 2.1 Affordances of Generative AI in Connection with Cybercriminal Activities

The growing tension over the misuse of GenAI models motivates examining these technologies through the lens of affordances theory (Gibson, 1979), as this theory considers both the capabilities of users and the objects they interact with. At its core, affordances theory posits that the design of objects and environments inherently suggests how they can be used, influencing human interaction and behavior (Gibson, 1979). When applied to the context of Information Systems, the affordance theory suggests that digital environments and tools provide individuals with cues about their potential uses and actions. In other words, affordances are not inherent in the person or technology; they emerge from their interaction.

Understanding these affordances is vital to recognizing GenAI models' extensive capabilities. By recombining their vast training datasets, GenAI can create content, such as code, video, text, images, and sound. However, when met with the malicious intentions of cybercriminals, the same capabilities of GenAI pose potential risks. When the capabilities of these AI models (as illustrated in the left circle of Figure 1) align with the goals of cybercriminals (as illustrated in the right circle of Figure 1), they lead to the emergence of undesirable affordances, as illustrated by the intersection of the two circles. These affordances enable attackers in several ways: they help them generate content, prepare programming code to embed in attack sites, aid in developing viruses and malware, and tailor attacks to specific targets.

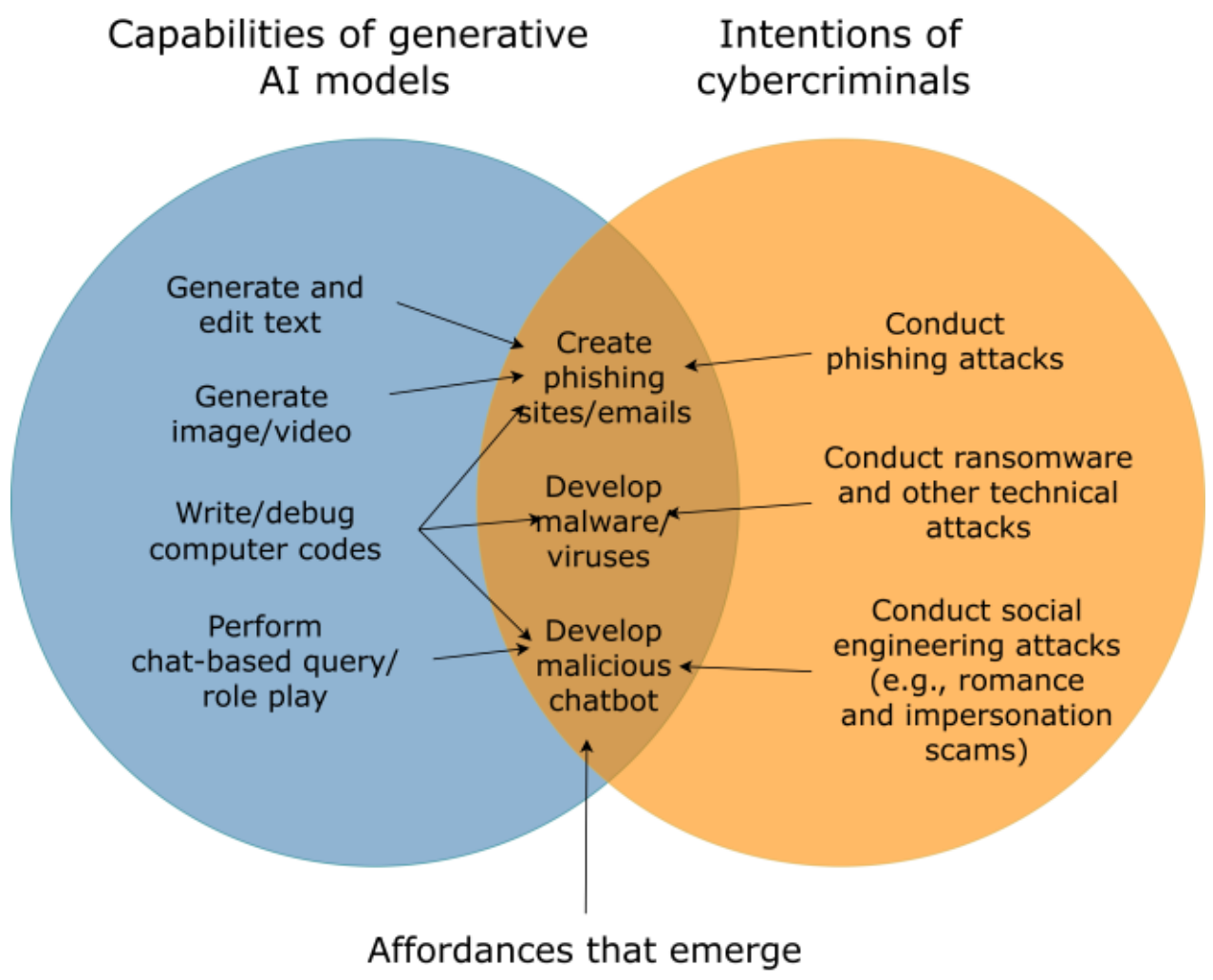

Figure 1: Examples of affordances that might emerge with the capabilities of generative AI models for cybercriminals

Next, we draw on the theory of technological amplification, which posits that while technology itself may be neutral, it often corresponds with an amplification of existing human intentions. This perspective is particularly relevant as it posits that GenAI, while ostensibly neutral, can amplify existing human intentions, whether beneficial or malicious.

### 2.2 Technological Amplification

Numerous technologies have had unintended consequences on society. For instance, cryptocurrencies have been perceived as both advantageous and detrimental. On the one hand, they offer decentralization, pseudonymity, and transparency, enhancing financial autonomy and privacy (Werbach, 2018), on the other hand, these same features can lead to concerns over security and regulation as criminals can use cryptocurrencies for dealing drugs, weapons trading, and other illegal activities.

Although technology may be inherently neutral, its application is tied to user intentions. This perspective is central to research on the integration of human factors and AI, where design choices can unintentionally facilitate the amplification of user behaviors and biases, necessitating careful socio-technical alignment (Chignell et al., 2023). Research has suggested that new technologies often amplify existing human intentions and that "*technology is merely a* *magnifier* [emphasis added] *of underlying human and institutional intent and capacity*" (Toyama, 2011, p. 77). This theoretical lens has been empirically applied to show how mobile phone services in rural Kenya, rather than universally uplifting users, amplified existing social inequalities by primarily benefiting the dominant network provider at the expense of marginalized, low-income women (Wyche et al., 2016). When introduced, the telephone, for example, did not alter existing social patterns. Instead, it amplified them by enhancing human communication efficiency, demonstrating how new technology can amplify existing predispositions—including negative ones—at an unprecedented scale. This is particularly evident when clear regulation, frameworks, and enforcement are absent. In the case of GenAI models, according to Bengio et al. (2024), the advancement of AI capabilities introduces significant risks, as AI systems have the potential to amplify social injustice, undermine social stability, support large-scale criminal activities, and enable automated warfare, targeted mass manipulation, and widespread surveillance. Additionally, the authors caution that these risks may be heightened, and new risks may emerge as companies work to develop autonomous AI agents, which are capable of using other information technology tools and service to interact with the world and pursue their objectives (Knight, 2025).

While responsible stakeholders explore how to govern AI for beneficial purposes, this process often takes years to get feedback, approval, and educate the public[1]. Meanwhile, bad actors such as cyber criminals with malicious intentions can exploit information technology without hesitation or known limitations.

[1] See, for example, how it took the EU seven years to create the General Data Protection Regulation: https://www.edps.europa.eu/data-protection/data-protection/legislation/history-general-data-protection-regulation_en

Overall, GenAI substantially enhances cybercriminal efficiency and effectiveness. For instance, these capabilities enable attackers to quickly develop sophisticated phishing sites, fraudulent websites designed to steal user credentials, with minimal effort or technical skills by generating the necessary images and authentic-looking content, such as blog posts or an institution's login and homepage. Additionally, with the ability to increasingly generate realistic human content, GenAI models could enable attackers to craft highly personalized phishing emails and websites, making them more effective (Roy et al., 2024; Schmitt & Flechais, 2024). Technical experts and law enforcement agencies have warned that hackers can use GenAI-enabled agents to engage in conversations with victims, facilitating social engineering attacks (FBI, 2024; Kosinski & Carruthers, 2025). With adaptive scams, GenAI models could even refine scams based on feedback, enhancing their persistence and adaptability in phishing campaigns and ultimately increasing their effectiveness. Additionally, with GenAI models, attackers can request help writing codes that automate processes such as mass phishing, blackmailing, and malware distribution. This reduces the time attackers spend per attack, amplifying underlying human intentions, especially when regulations and enforcement are lacking.

## 3. Empirical illustration: Interrupted time series analysis

### 3.1. Method

To empirically illustrate the plausibility of the affordance plus amplification framework, we analyze time series data from two cybercrime reporting platforms. The goal is to assess whether changes in reported malicious activities around the public release of ChatGPT 3.0 are consistent with our theoretical expectations. We analyze the AbuseIPDB and Chainabuse datasets before and after this event. Interrupted time series analysis (ITSA) is a suitable and robust method for evaluating changes in a time series following an intervention, particularly when randomized control trials are not possible. By comparing the pre-intervention trend with the post-intervention trend, we can assess whether there is a significant change in the slope of the cybercrime amount associated with the launch of GenAI models while controlling for underlying trends that could confound the results (Bernal et al., 2017). Furthermore, ITSA allows for evaluating the immediate effects and the long-term trends post-intervention, which is essential in understanding the dynamic nature of cybercrime activities concerning technological advancements (Bernal et al., 2017). This analysis can be enhanced by adding a control series to the model to rule out alternative explanations for observed changes, increasing the study's internal validity.

One of our key investigative parameters is the comparison of the frequency of reported scams before and after the introduction of generative AI models. In our analysis, we used the date (November 30, 2022) that ChatGPT 3.0, the first commercialized generative AI model, was released to the public as the intervention date. We treat data before the release date of ChatGPT as pre-intervention and data after the release as post-intervention using the following equation:

$$Y_t = \beta_0 + \beta_1 \times Time_t + \beta_2 \times Intervention_t + \beta_3 \times Time\ after\ intervention_t + \varepsilon_t$$

$Y_t$ is the dependent variable representing the scam/malicious activity counts at time t. $Time_t$ is the continuous time variable representing the number of time units since the beginning of the study period. $Intervention_t$ is a dummy variable indicating the presence of the intervention (0 before the intervention and 1 after the intervention). $Time\ after\ intervention_t$ is a continuous variable representing the time since the intervention occurred, starting from 0 and increasing with time after the intervention. $\beta_0$ is the intercept, $\beta_1, \beta_2, and\ \beta_3$ are the coefficients for each of the respective terms in the model and $\varepsilon_t$ is the error term at time t.

By analyzing the coefficient $\beta_2$, we can measure the immediate shift in the level of scam/malicious activity counts that occurred at the time of the intervention. $\beta_3$ helps understand the trend of the scam/malicious activity counts following the intervention. This variable is helpful in assessing the long-term effects of an intervention. It can show whether the impact of the intervention strengthens, weakens, or remains constant over time (Wagner et al., 2002). Prior to modeling, we tested the data for stationarity using the Augmented Dickey-Fuller (ADF) test, which checks for a unit root. We use an alpha level of 5%; if the p-value is less than 0.05, indicating stationarity, no differencing was applied (d = 0); otherwise, first-order differencing was used (d = 1) (Asteriou & Hall, 2011).

### 3.2.Data Analysis and Results

Evaluating both general IP abuse data and specific crypto-related scam data allowed us to better understand the nuances of GenAI's effects and increase the robustness of our findings. These platforms have emerged as pivotal resources in combating cybercrime, providing extensive data on malicious online activities (Huang et al., 2023; Lewis et al., 2020).

#### 3.2.1. Study 01: AbuseIPDB.com Data

*Data Collection of Study 01*

AbuseIPDB is a website dedicated to combating the proliferation of hackers, spammers, and other harmful online activities. Its mission is to bolster internet security by maintaining a comprehensive blacklist that webmasters and system administrators rely on. In many organizations, this resource has become a crucial component of the IT cyber-response toolkit (Cisco, 2025). Specialized users, such as IT firms and server managers, can access its API or integrate their security systems with AbuseIPDB (Bassey, 2022; Maltego, 2021). Researchers, particularly those using its data to train machine learning models in cybersecurity research (Lewis et al., 2020), also benefit from this service. It regularly updates its data but only displays reports from the most recent 30-day period, restricting direct access to more extensive historical data. To overcome this limitation and build a time series dataset, we used the Internet Archive's Wayback Machine (web.archive.org) to circumvent the limited data display on the live website. This approach allowed us to systematically access and extract daily archived snapshots of publicly reported IP addresses. Through this effort, we gathered data spanning from January 1, 2022, approximately one year before the launch of ChatGPT, to October 29, 2023, one year after the public release of ChatGPT 3.0. As a result, we compiled a dataset of 464,190,074 malicious IP addresses reported during that period.

*Results of Study 01*

We first conducted an ADF test to examine the stationarity of the data. The results yielded a p-value < 0.001, thereby confirming the stationarity. We then proceed to the model estimations. We treat November 30, 2022, the public release of ChatGPT, as a salient 'access shock' that rapidly expanded broad public exposure to a high-capability, conversational GenAI system. Table 1 reports the results of the model using the weekly counts. The model $R^2$ indicates that the model can explain 58.5% of the variation in the amount of cybercrime. Key coefficients within the model include *Time, Intervention,* and *Time after intervention.* The coefficient for *Time* is positive ($\beta_1$= 14,110), signifying a general increasing trend in malicious activities over time, independent of the intervention.

Table 1: Interrupted time series analyses on the number of malicious IP addresses from AbuseIPDB.com

| Estimate | Coefficient | Std. Error | t-value | p-value | 95% CI |
|---|---|---|---|---|---|
| Constant ($\beta_0$) | 3,931,000 | 179,000 | 21.981 | <0.001 | [3,580,000, 4,290,000] |
| Time ($\beta_1$) | 14,110 | 7,331.060 | 1.925 | 0.057 | [-454.466, 28,700] |
| Intervention ($\beta_2$) | 1,120,000 | 249,000 | 4.499 | <0.001 | [625,000, 1,610,000] |
| Time after intervention ($\beta_3$) | -16,670 | 9,376.785 | -1.778 | 0.079 | [-35,300, 1,960.543] |
| $R^2$ = 58.5%; adjusted $R^2$= 57.1%; DV: Number of weekly reported IP addresses | | | | | |

The *Intervention* coefficient, which represents the introduction of GenAI models, is positive and statistically significant ($\beta_2$=1,120,000; p-value <0.001), indicating a substantial increase in cybercrime following this introduction. These results suggest that the public release of generative AI correlates with an immediate increase in cybercrime. The results of study 01 indicate that the positive *Intervention* coefficient ($\beta_2$) in the model indicates that the introduction of generative AI models correlates with an overall increase in the amount of cybercrime. Regarding the long-term effect (*Time after intervention,* $\beta_3$), the model suggests a slight post-intervention decrease of 16,670 cases in cybercrime following the immediate spike of 1,120,000 cases; however, the slight decrease is not statistically significant.

### 3.2.2. Study 02: Chainabuse.com Data

*Data Collection of Study 02*

For this study, we collected data from Chainabuse.com, a public reporting platform for malicious cryptocurrency-related activities designed to enhance the safety of Web 3.0 ecosystems. The platform allows users to report fraudulent incidents and collaborates with leading organizations to expedite investigations. To compile a comprehensive dataset, we developed and deployed a custom Python script to collect all publicly accessible archival reports from the website. Through this, we aimed to gauge the overarching influence of GenAI on the cybercrime landscape in the context of cryptocurrency. Indeed, financial innovation and technological advances are enabling criminals to exploit security deficiencies in cryptocurrency markets to launder illicit funds or facilitate fraud at unprecedented rates (Akartuna et al., 2022).

The dataset encompassed records from January 1, 2021, to October 10, 2023, with 281,115 reports of crypto-related scams. Similar to study 01, we divided the data into two segments, pre-intervention and post-intervention.

*Results of Study 02*

The ADF test results on the data show strong stationarity, with an ADF statistic of p-value < 0.001. These results indicate that the data is stationary without requiring differencing or transformations (Asteriou & Hall, 2011). Our analysis of weekly scam data, reported in Table 2, reveals a good model fit, indicating that the model explains about 67.2% of the variation in weekly scam amount. The *Time* variable shows a minimal increasing amount in scam reports over time, with a coefficient, $\beta_1$= 4.214 (p-value = 0.005), suggesting an increase of approximately four cases per week.

Table 2: Interrupted time series analyses on the crypto-related cybercrime from Chainabuse.com

| Estimate | Coefficient | Std. Error | t-value | p-value | 95% CI |
|---|---|---|---|---|---|
| Constant ($\beta_0$) | 528.871 | 85.427 | 6.191 | <0.001 | [359.999, 697.743] |
| Time ($\beta_1$) | 4.214 | 1.491 | 2.827 | 0.005 | [1.267, 7.161] |
| Intervention ($\beta_2$) | 721.673 | 154.720 | 4.664 | <0.001 | [415.821, 1027.525] |
| Time after intervention ($\beta_3$) | 10.178 | 5.006 | 2.033 | 0.044 | [0.281, 20.075] |

$R^2$ = 67.2%; adjusted $R^2$=66.5%; DV: Number of weekly reported scams

Notably, the *Intervention* variable has a significant impact, with a coefficient $\beta_2$= 721.673 (p-value < 0.001), indicating a considerable and immediate increase in the level of overall scam reports following the public release of ChatGPT3.0. Interestingly, the *Time after intervention* variable, with coefficient $\beta_3$= 10.178 (p-value = 0.044), suggests a further long-term increase in cryptocurrency-related scam reports over time, post-intervention. In summary, the model indicates a substantial increase in reported scams following an intervention.

### 3.2.3. Robustness Checks

We conducted robustness checks on the models to ensure the validity of our findings. ARIMA models account for both autocorrelation and moving average components in time series data, making them well-suited for handling the temporal dependencies seen in the counts of malicious IP addresses or crypto-related scams. By incorporating exogenous variables (e.g., the intervention date), ARIMA offers a robust complementary check on the main interrupted time series results, reinforcing the reliability of the findings (Shumway et al., 2017).

Furthermore, in the cryptocurrency realm, a surge in the exchange rate between cryptocurrencies and the USD might influence the amount of cybercrime due to the heightened motivation of cybercriminals to execute attacks as cryptocurrency values rise and more users are drawn into cryptocurrencies' speculative nature, potentially driving up crime amount. Bitcoin is a good proxy for this purpose because it is one of the earliest cryptocurrencies and has the largest market cap, accounting for approximately 60% of the total cryptocurrency market cap (CoinMarketCap, 2025). After controlling for Bitcoin prices in Study 2, the findings continue to be statistically significant in the ARIMA models. This reinforces the relationship between the public release of generative AI models on cybercrime. Detailed outcomes from these alternative models are reported in Appendix Tables A1 and A2.

## 4. Discussion

This paper first offers a plausible explanation, through a dual theoretical lens, of how GenAI can be used for cybercrime, and then provides empirical evidence that speaks to concerns about GenAI's dual-use nature by showing its association with increased malicious activities and cryptocurrency scams. Study 1, based on AbuseIPDB data, demonstrates a significant increase in the number of reported malicious IP addresses following the public availability of GenAI models. Study 2, based on Chainabuse data, shows both an immediate jump and a sustained post-intervention increase in reported cryptocurrency scams. In this section we integrate these patterns with our affordance and technological amplification framework and clarify what they imply for how generative AI is reshaping cyber risk.

From an affordance perspective, the release of GenAI actualized new 'action possibilities' for malicious actors, such as automated content generation and low-code scripting. Concurrently, from a technological amplification perspective, our results align with the theory that these new affordances may serve to magnify existing criminal intent on a massive scale once the tools become broadly accessible. While an immediate increase in the number of cybercrime cases was observed, the amount decreased, albeit insignificantly, post-intervention, adding nuance to the relationship between generative AI and cybercrime. One plausible interpretation is a socio-technical adaptation effect: as awareness of AI-enabled deception increased, platforms, firms, and security vendors may have adjusted detection rules, user warnings, and authentication controls, partially offsetting the initial surge in reported abuse. Large organizational evaluations likewise find that anti phishing programs can shift both knowledge and protective behavior when the content is adapted to user characteristics (Sumner et al., 2022). Additionally, security systems such as security information and event management (SIEM), intrusion detection systems (IDS), and intrusion prevention systems (IPS) are becoming more advanced (Li et al., 2019). Some even implement advanced AI models to detect nefarious activities with the assistance of GenAI.

In Study 02, the cryptocurrency-related scam data also revealed a significant increase in reported scams following the public release of the first GenAI model.

More interestingly, the long-term effect after the intervention indicates a sustained increase in cryptocurrency-related scams. This differentiation between context (whether cybercrime more broadly or crypto-related) underscores the need for a more refined understanding of GenAI’s impact on different types of cybercrime. The findings emphasize the significance of examining cybercrime categories when assessing the impact of GenAI and suggest the need for tailored strategies to combat cyber threats effectively based on the type of crime or attack.

It is also critical to consider an alternative interpretation for the immediate spike observed in our data. The public release of ChatGPT was a major global media event that significantly increased discourse around AI's capabilities, both good and bad. Accordingly, we interpret our estimates as shifts in reporting time-series consistent with increased abuse, increased vigilance, or both; future work can separate these channels by incorporating exposure proxies (for example, media-attention indices or platform policy-change timelines) and by adding control series that are unlikely to be affected by GenAI misuse.

Our findings empirically substantiate our theoretical framework, which is rooted in human-computer interaction: the interplay between GenAI's capabilities and cybercriminals' goals gives rise to unwanted affordances (e.g., for creating malware or phishing sites). The sharp, immediate increase in malicious reports shown in our ITSA models provides empirical weight to this theoretical claim, suggesting that these new 'action possibilities' were not only perceived but rapidly exploited by malicious actors, which is consistent with the theory of technological amplification regarding the efficiency and scale of their activities. Next, we discuss the theoretical and practical implications of this research.

### 4.1. Implications for Research

The results suggest three practical governance control points: (1) model developers, through pre-deployment misuse-resistance testing and ongoing capability evaluation; (2) hosting and distribution platforms, through risk tiering, model vetting, and fast takedown processes for explicitly abuse-oriented artifacts; and (3) regulators, through auditable obligations for high-risk AI systems and incident-reporting requirements that enable longitudinal monitoring of downstream harms.

First, while some major tech companies are considering cutting their responsible AI staff, raising public concerns (Murgia & Criddle, 2023), our findings emphasize the crucial need for ethical and responsible AI considerations in IS research (Gregor, 2024). Research examining public trust in AI-based systems has demonstrated that concerns related to AI misuse, ethical considerations, and contextual factors significantly influence trust levels, underscoring the importance of embedding ethical frameworks into AI development and deployment (Omrani et al., 2022). Although AI developers and companies have taken steps to prevent the misuse of their technologies in cybercrime by implementing safeguards and ensuring responsible, accountable AI implementations at both individual and institutional levels, the current response from many AI companies remains hard-coded to prevent unethical behavior (Mok, 2023). For instance, X blocked certain Taylor Swift searches to prevent the spread of deepfake explicit photos on the internet (PBS, 2024).

These examples show that traditional approaches are still being used to tackle increasingly sophisticated problems. Given the affordances of GenAI, which allow for decentralized and uncensored model training and deployment, and the amplification of malicious intent through these technologies, research on ethics and responsible AI must be continuously updated to address emerging challenges. Studies have shown that traditional IS ethics principles differ from AI ethics principles (Mirbabaie et al., 2022).

It is imperative for academics and practitioners to collaborate on adaptive frameworks that proactively address these emerging ethical and safety challenges in AI systems. Current initiatives like the unified framework of five principles for AI in society (Floridi & Cowls, 2022) or the ethics of artificial intelligence (Unesco, 2025) demonstrate the potential of such collaboration to create flexible, proactive solutions for ethical AI development.

Second, it is critical to examine how the decentralization of model training and deployment challenges established governance mechanisms. At the time of writing, large-scale model repositories host extensive numbers of user-trained and customized models, lowering the cost of deploying specialized GenAI systems outside traditional organizational controls. This diffusion shifts governance from single-firm compliance toward ecosystem governance, where repositories, cloud providers, and downstream integrators become key points of accountability. While much of the AI governance research focuses on the responsible use of AI by institutions to comply with current laws and regulations and to protect their consumers from vulnerabilities (Papagiannidis et al., 2025), these governance mechanisms do not seem to be effective for open-source and private models deployed outside institutions. In other words, much more research needs to be conducted to address the decentralization of technology outside of corporate practices. For instance, cybercriminals might be using their homegrown AI tools (e.g., private LLMs) to conduct their operations. Many of these models can answer a range of uncensored questions, such as bypassing security measures, creating destructive weapons, or generating explicit content. The affordances of these decentralized models, combined with the amplification of malicious intent, pose significant risks that current governance structures are ill-equipped to handle. The unintended negative effects of AI technologies, including their potential to magnify inequalities and exacerbate existing vulnerabilities. This trend necessitates transformative governance solutions that prioritize ethical considerations and fairness in AI design, deployment, and ownership (Bircan & Özbilgin, 2025). Empirical evaluations of GenAI systems on certified ethical hacking tasks further underscore how tools like ChatGPT and Bard can already perform complex offensive security activities, reinforcing the need for updated governance and assessment frameworks (Raman et al., 2024).

Because of their open-source and customizable nature, the guardrails on these technologies seem ineffective at this moment. Additionally, at the organizational level, cybersecurity strategies, such as secure-by-design principles, are being implemented, though only 24% of generative AI initiatives are currently secured (Lindemulder & Kosinski, 2025). Coupled with our interrupted-time-series evidence of an immediate post-ChatGPT surge of more than 1.1 million malicious IP reports and a sustained rise in crypto-related scams, this shortfall underscores the need for more research. Based on this, we propose three potential research directions:

(1) Despite incremental hard-coded defenses, recent prompt-injection incidents reveal that both institution-hosted and customized LLMs can still be "jail-broken," enabling attackers to weaponize the very affordances these models provide. From a human-computer interaction perspective, 'jailbreaking' is a form of adversarial use where users probe and exploit system limitations to bypass intended safeguards. As this is a key challenge in human-computer interaction, we call for future research to look into these adversarial interactions to understand the patterns of user behavior that lead to security breaches and to inform the design of more robust, adaptive safety measures. These concerns are consistent with broader systematizations of offensive AI, which map how adversaries can integrate AI capabilities across the cyber kill chain to increase the scale and sophistication of attacks (Mirsky et al., 2023).

(2) With over one million models already available on Hugging Face, and virtually no uniform oversight, the locus of governance is shifting from developers to model-hosting platforms. The success of platforms often depends on creating an environment that is welcoming to contributions that are perceived to be of high quality (Tsai & Pai, 2021) As this perception grows, platforms build a positive reputation that it will lead to others recommending its use, thus further increasing its popularity and success (Chu & Manchanda, 2016). However, if the model-hosting platforms gain a reputation for hosting models known to be of benefit for cybercrime, it might threaten the success of these platforms. We call for socio-technical research into governance mechanisms for model-hosting platforms. How can these platforms design interfaces, policies, and community moderation systems that disincentivize misuse while preserving the creative and innovative potential of open-source development?

(3) Our findings illustrate how GenAI affordances amplify existing criminal intent, suggesting that purely technical fixes will be insufficient unless aligned with emerging regulatory frameworks such as the EU AI Act and organizational risk-management practices. These frameworks need to embrace a multi-layered strategy that includes fast-acting interventions, alongside longer-term policies that reflect balanced thinking to correct short-term learning. More research is needed on the best way to implement these kinds of multi-layered strategies that combine technical, regulatory, and organizational interventions to effectively address the risks posed by easily deployable LLMs in open-source ecosystems. As the pace of information technology development only intensifies, this kind of strategic response will be imperative to protect individuals and embrace innovation.

## 4.2. Implications for Practice

First, the rise in cybercrime-related behaviors post-GenAI underscores the pressing need to enhance public awareness and training about the risks of these new technologies (Zhuang et al., 2020). Therefore, novel and tailored educational initiatives and training designs, such as gamified interactive storytelling (e.g., Dincelli & Chengalur-Smith, 2020) or antineutralization technique (e.g., Barlow et al., 2018) may be necessary to equip end-users with the skills to identify and protect themselves from AI-assisted phishing, social engineering using deepfakes, and other scams; moreover, the breadth, depth, and finesse of organizational knowledge mechanisms shape how employees construe security threats and adopt secure behaviors (Mady et al., 2023). Beyond message content and format, recent work shows that the messenger's perceived characteristics (e.g., authority or expertise) significantly influence compliance intentions to security messages (Schuetz et al., 2025).

Complementing these designs, evidence shows that raising general cybersecurity and policy awareness measurably shifts employees' compliance attitudes and strengthens supply-chain reactive capabilities, especially in small and medium enterprises (Wong et al., 2022). Individuals' legitimacy judgments of cybersecurity initiatives strongly shape whether they comply with security policies (Cram & D'Arcy, 2023) and encouraging proactive information security behaviors can complement training by motivating self-initiated, change-oriented protection efforts (Xu et al., 2024). Although broad AI literacy helps, training should prioritize vulnerable populations such as children, older adults, and cryptocurrency adopters who are frequently targeted. At the same time, younger digital-native users can appropriate GenAI for opportunistic or unethical purposes, which creates a parallel insider-style risk that organizations must anticipate (Smith et al., 2025). Given offenders' growing toolset, systematic detection technologies can further mitigate harm by helping potential victims recognize and report scams.

Second, cybersecurity strategies must be dynamic and adaptable to effectively counter the ever-changing tactics employed by cybercriminals leveraging AI technologies. This necessitates ongoing monitoring of cybercrime trends and timely updating of threat intelligence repositories (Wagner et al., 2019). This also echoes recent forecasting-oriented work that links cyber threat anticipation with the identification of important mitigation technologies (Almahmoud et al., 2025). The varying impact of generative AI on different scam types suggests a need for tailored approaches to combat each type effectively, including developing AI models that counteract AI-generated cyber threats and exploring new defense mechanisms. Our findings also support the need for updated policies to address the challenges posed by generative AI, particularly in cybersecurity and privacy. Threat-modeling approaches developed for sensitive domains such as reproductive healthcare privacy offer structured methods for anticipating and mitigating emergent AI-driven risks (McDonald et al., 2025). Specifically, with the increasing ability of users to train and deploy open-source AI models locally using local hardware or cloud computing, institutional safeguard mechanisms may not apply to these open-sourced and privately trained and deployed models. This highlights a pressing need for research into scenarios where institutions are not central to AI governance.

Third, while generative AI models are inherently neutral, it is crucial to recognize that, through affordances and technology amplification perspectives, human usage can determine the direction of the outcomes. The critical role of institutional and corporate power in shaping these outcomes cannot be overlooked; just as a mobile network provider's design and business decisions can amplify economic disparities in developing regions (Wyche et al., 2016), the decisions made by AI developers and platform hosts will significantly influence whether their technologies primarily amplify beneficial or malicious human intent.

While non-legally binding AI governance frameworks such as those from Microsoft (i.e., AI privacy and security, reliability, safety, fairness, inclusiveness, transparency, and accountability) (Microsoft, 2025) or the NIST AI Risk Management Framework (i.e., map, measure, manage, and govern) (NIST, 2025) might work well for institution-based models to design, develop, deploy, use, and govern AI, but it might not be effective with open-sourced or uncensored models deployed locally or offshore by malicious actors. Along with its benefits, the decentralization of AI models could pose greater risks to humanity due to the rapid pace of development, as both hardware and algorithms continue to improve. For instance, AI computing chips have become 1.4 times more cost-effective each year, while AI training algorithms have grown 2.5 times more efficient in the same period (Bengio et al., 2024).

Hence, besides managing the tools, stakeholders must also address the human behaviors and expectations that shape their application (Minkkinen et al., 2023). Therefore, in terms of managing AI, in addition to managing the tools themselves via guidelines and frameworks, legally binding mechanisms need to be put in place to mitigate risks. For instance, the European Union (EU) passed its AI Act in 2024 to focus on managing risks by classifying AI systems into categories of unacceptable risk, high risk, limited risk, and minimal risk (European Union, 2025). This approach implies that it might not be possible to apply definitive guardrails directly to the tool. Instead, the focus has shifted to the tools' implications for their users.

Overall, as the boundaries between beneficial innovation and malicious exploitation become increasingly blurred by open-sourced or uncensored AI models, we must ask ourselves: how can we safeguard society from emerging AI threats without stifling the transformative potential that these technologies hold? This tension between regulation and innovation may well define the trajectory of AI adoption and its broader social implications in the years to come.

### 4.3.Limitations

While this paper employs a multi-study approach to assess a theory-driven empirical analyses, we recognize the value of alternative strategies in future research. Specifically, conducting qualitative research through in-depth interviews with white-hat hackers and system/network administrators could provide deeper insights into cybercriminals' methods to leverage AI for crime. Such an understanding has the potential to inform the development of more practical, action-oriented strategies to combat cybercrime. Additionally, even though interrupted time series analyses can demonstrate significant changes before and after the public availability of these models, causality must be interpreted with caution. Furthermore, our data relies on public reporting platforms, and the observed trends could be influenced by changes in reporting behavior.

## 5. Conclusion

Drawing on affordance theory and technological amplification, this paper provided empirical evidence that GenAI's democratization is associated with escalating cybercrime-related behaviors. This work corroborated concerns about AI's dual-use potential, illustrating how its capabilities and attacker motivation can increase malicious activities. These findings underscore the urgent need for AI developers, cybersecurity stakeholders, and society to address the heightened risks posed by this technology.

## Appendix

Robustness check using alternative statistical models.

Table A1: Analysis of the malicious activities from AbuseIPDB.com (study 01), using autoregressive integrated moving average

| Estimate | Coefficient | Std. Error | z-value | p-value | 95% CI |
|---|---|---|---|---|---|
| Constant ($\beta_0$) | 3,931,000 | 459,000 | 8.591 | <0.001 | [3,030,000, 4,830,000] |
| Time ($\beta_1$) | 14,110 | 15,700 | 0.921 | 0.357 | [−16,600, 44,900] |
| Intervention ($\beta_2$) | 1,120,000 | 307,000 | 3.647 | <0.001 | [518,000, 1,720,000] |
| Time after intervention ($\beta_3$) | −16,670 | 20,300 | -0.835 | 0.404 | [-56,600, 23,200] |
| ar.L1 | 0.6032 | 0.070 | 4.917 | <0.001 | [0.522, 0.796 |
| Sigma2 | 1.918e+11 | 0.789 | 2.43e+11 | <0.001 | [1.92e+11, 1.92e+11 |

DV: Number of weekly reported IP addresses; ARIMA(1,0,0) was selected based on the Bayesian Information Criterion (BIC), Akaike Information Criterion (AIC), and stationarity tests.

Table A2: Analysis of the crypto-related cybercrime from Chainabuse.com (study 02), controlling for Bitcoin price using using autoregressive integrated moving average

| Estimate | Coefficient | Std. Error | z-value | p-value | 95% CI |
|---|---|---|---|---|---|
| Constant ($\beta_0$) | 870.2319 | 636.060 | 1.368 | 0.171 | [-376.423, 2116.887] |
| Time ($\beta_1$) | 2.2582 | 5.363 | 0.421 | 0.674 | [-8.25, 12.769] |
| Intervention ($\beta_2$) | 700.1294 | 327.724 | 2.136 | 0.033 | [57.802, 1342.457] |
| Time after intervention ($\beta_3$) | 13.5944 | 12.138 | 1.120 | 0.263 | [-10.196, 37.385] |
| Bitcoin price ($\beta_4$) | -0.0065 | 0.010 | -0.629 | 0.529 | [-0.02, 0.014] |
| ar.L1 | 0.6462 | 0.087 | 7.420 | <0.001 | [0.475, 0.817] |
| Sigma2 | 1.136e+05 | 1.09e+04 | 10.388 | <0.001 | [9.22e+04, 1.35e+05] |

DV: Number of weekly reported crypto-related scams; ARIMA(1,0,0) was selected based on the Bayesian Information Criterion (BIC), Akaike Information Criterion (AIC), and stationarity tests.